\documentclass[default,iicol]{sn-jnl}

\usepackage{natbib}
\setcitestyle{authoryear,open={(},close={)}} 
\PassOptionsToPackage{warn}{textcomp}  
\usepackage{textcomp}                  
\usepackage{mathptmx}                  
\usepackage{times}                     
\usepackage{cite}                      
\usepackage{tabu}                      
\usepackage{booktabs}                 

\usepackage{enumitem}
\usepackage{microtype}
\usepackage{setspace}
\usepackage{siunitx}
\usepackage{graphicx}
\usepackage[misc]{ifsym}

\jyear{2021}%

\theoremstyle{thmstyleone}%
\theoremstyle{thmstyletwo}%

\theoremstyle{thmstylethree}%

\begin{document}

\title[Article Title]{The Effects of Spatial Configuration on Relative Translation Gain Thresholds in Redirected Walking}



\author[1]{\fnm{Dooyoung} \sur{Kim}}\email{dooyoung.kim@kaist.ac.kr}

\author[1]{\fnm{Seonji} \sur{Kim}}\email{seonji.kim@kaist.ac.kr}

\author[2]{\fnm{Jae-eun} \sur{Shin}}\email{jaeeunshin@kaist.ac.kr}

\author[1]{\fnm{Boram} \sur{Yoon}}\email{boram.yoon1206@kaist.ac.kr}

\author[3]{\fnm{Jinwook} \sur{Kim}}\email{jinwook.kim31@kaist.ac.kr}

\author[3]{\fnm{Jeongmi} \sur{Lee}}\email{jeongmi@kaist.ac.kr}

\author*[1,2]{\fnm{Woontack} \sur{Woo}}\email{wwoo@kaist.ac.kr}

\affil*[1]{\orgname{KAIST  UVR  Lab}, \orgaddress{\street{291, Daehak-ro, Yuseong-gu}, \city{Daejeon}, \postcode{34141}, \country{Republic of Korea}}}

\affil[2]{\orgname{KAIST KI-ITC Augmented Reality Research Center}, \orgaddress{\street{291, Daehak-ro, Yuseong-gu}, \city{Daejeon}, \postcode{34141}, \country{Republic of Korea}}}

\affil[3]{\orgname{KAIST Visual Cognition Lab}, \orgaddress{\street{291, Daehak-ro, Yuseong-gu}, \city{Daejeon}, \postcode{34141}, \country{Republic of Korea}}}

\abstract{
In this study, we explore how spatial configurations can be reflected in determining the threshold range of Relative Translation Gains (RTGs), a translation gain-based Redirected Walking (RDW) technique that scales the user's movement in Virtual Reality (VR) in different ratios for width and depth. While previous works have shown that various cognitive factors or individual differences influence the RDW threshold, constructive studies investigating the impact of the environmental composition on the RDW threshold with regard to the user's visual perception were lacking. Therefore, we examined the effect of spatial configurations on the RTG threshold by analyzing the participant's responses and gaze distribution data in two user studies. The first study concerned the size of the virtual room and the existence of objects within it, and the second study focused on the combined impact of room size and the spatial layout. Our results show that three compositions of spatial configuration (size, object existence, spatial layout) significantly affect the RTG threshold range. Based on our findings, we proposed virtual space rescaling guidelines to increase the range of adjustable movable space with RTGs for developers: placing distractors in the room, setting the perceived movable space to be larger than the adjusted movable space if it's an empty room, and avoid placing objects together as centered layout. Our findings can be used to adaptively rescale VR users' space according to the target virtual space's configuration with a unified coordinate system that enables the utilization of physical objects in a virtual scene.}


\maketitle
\section{Introduction}
\label{sec:1}
Since COVID-19, remote collaboration has become commonplace, and an avatar-based telepresence system using Virtual Reality (VR) and Augmented Reality (AR) Head-Mounted Displays (HMDs) is drawing attention as a means to provide immersion and co-presence between users~\citep{orts2016holoportation, piumsomboon2018mini, kim2021multi}. In most cases, the virtual space accessed by the VR client is more expansive and different from the tracked space in which the user physically exists. To overcome this limitation, Redirected Walking (RDW), which transforms users' locomotion in VR to a level that users do not notice, was proposed~\citep{razzaque2005redirected}. Most RDW techniques utilize rotation gain, which can achieve a more significant effect on expanding explorable space than the translation gain~\citep{dong2020dynamic,strauss2020steering, williams2021arc}. Nevertheless, it is challenging to utilize objects and walls in real space with rotation-based RDW on account of the distortion of coordinate systems between real and virtual spaces. On the other hand, the translation gain-based RDW can generate a unified coordinate system between the virtual space and the tracked space while rescaling the VR space. Relative Translation Gains (RTGs), a translation gain-based RDW that transforms the horizontal and vertical ratios of space differently, have been proposed to transform virtual spaces while maintaining a unified coordinate system~\citep{kim2021adjusting}.

As RDW technologies transform the users' perceived locomotion in VR, it is necessary to modify redirecting values within the threshold range where users do not significantly notice the difference in their movement~\citep{steinicke2009estimation, grechkin2016revisiting}. Since RDW technology visually changes the users' movement, the threshold range varies depending on the subjects' visual cognition factors in a virtual environment. Previous studies confirmed that subjects underestimate their speed in a virtual environment~\citep{ash2013vection,banton2005perception,bruder2011tuning,caramenti2018matching}. Moreover, users' individual differences and the presence of cognitive tasks while they are walking also affect the RDW threshold~\citep{williams2019estimation,bruder2015cognitive,rietzler2018rethinking,sakono2021redirected}. Although the purpose of RDW is to match a physical tracked space with a virtual target space, there was a lack of research systematically summarizing how RDW thresholds are affected by the configuration of virtual spaces.

In this study, we analyzed users' responses about their perceived speed and gaze information to reveal how the RTG threshold is affected by the configuration of virtual space. Based on our previous work on the effect of size and objects on RTG threshold~\citep{kim2022effects} (Study 1), we conducted an additional study (Study 2) focusing on the spatial layout of the virtual space to expand upon the implications of its findings and conduct an integrated analysis on how spatial configurations affect the range of RTG thresholds. In Study 1, RTG thresholds were estimated under a total of six spatial conditions depending on the size of the virtual space and the presence or absence of objects (3 $\times$ 2) which were configured by differing Angles of Declination (AoDs) between the user's gaze and the virtual horizon. In Study 2, we estimated the RTG threshold under eight spatial conditions according to the size of the virtual space and the spatial layout (2 $\times$ 4) and analyzed the changes in gaze according to the virtual space configuration through AoD distribution and gaze heatmap.

Our results confirmed that all three components of spatial configuration (size, object presence, spatial layout) significantly affected the RTG threshold range. In Study 1, the RTG threshold range was greater in the largest size condition than in two smaller size conditions, and the RTG threshold range further increased in the furnished room than in the empty one. In Study 2, the RTG threshold range was greater in the scattered layout than in the centered layout. In connection with the results of examining AoD data, we confirmed that the AoD distribution decreased as the size of the empty room increased. Moreover, even though the placement of objects lowered the AoD distribution, the effect of the object as a distractor had a more significant effect on the RTG threshold range than the size of the virtual space. The gaze heatmap examined in Study 2 shows that when objects are gathered into a single cluster, they act as a guide to help subjects recognize their perceived speed in VR, leading to the RTG threshold range being lower in the centered layout than in the empty virtual room.

Through this study, we investigated the effect of the spatial configuration of VR room on translation-based RDW threshold. In addition, the user's gaze data and the subject's response were analyzed together to more quantitatively analyze the tendency of the user's gaze distribution to vary depending on the configuration of the virtual space and the effect on the RTG threshold. Our findings could be used to adaptively expand and transform VR users' tracked space according to the spatial configuration of the target virtual environment. We derived three virtual space rescaling guidelines for increasing the range of adjustable movable space with the RTG by synthesizing our results. First of all, placing disctractors in the virtual room to increase the adjustable movable space. Second, for empty spaces, set the perceived movable space to be larger than the adjusted movable space. Lastly, avoid placing objects where they are constantly in a user's sight. Through our discovery, developers could set a suitable range of RTG according to the spatial configuration of the virtual room and rescale the VR user's space to fit the target virtual space, adaptively. Furthermore, the space modification with RTG could maintain the coordinate system between physical and virtual space, so it can also be used for XR remote collaboration where AR/VR users from heterogeneous spaces are co-located in a mutual space.

\section{Background}
\label{sec:2}

\subsection{Redirected Walking}

Redirected Walking (RDW) is a VR locomotion technique that allows users to explore a more expansive virtual space with natural walking from the limited tracked space of VR users~\citep{razzaque2005redirected,ropelato2021hyper,nilsson2018natural}. RDW utilizes three types of gain: rotation gain that deforms the angle, translation gain that deforms the speed, and curvature gain that deforms through applying offset. Various RDW controllers have been suggested to utilize these RDW gains to allow users to walk around an ample virtual space. The steer-to-center (S2C) method directs the user to the center of the tracked space, steer-to-orbit (S2O) that redirects the user to a specific orbit, and steer-to-multi-target (S2OT) that directs to multiple targets were frequently used as baseline RDW controllers~\citep{hodgson2013comparing,chen2018redirected}. In addition to directing users with such a fixed target, Strauss et al. proposed RDW with reinforcement learning~\citep{strauss2020steering,chen2021reinforcement}, and Williams et al. suggested space-adaptive RDW controllers by comparing the layout of real and virtual spaces in real-time~\citep{williams2021arc,williams2021redirected}. Dong et al.~\citep{dong2017smooth} introduced Smooth Assembly Mapping (SAM) to decompose ample virtual space and assemble them to the tracked space. Moreover, RDW controllers for multiple users were proposed to enable collaborative scenarios with RDW~\citep{lee2020optimal,dong2020dynamic}. Recently, RDW libraries that can compare each performance of RDW controllers were also presented to strive for practical use of RDW~\citep{azmandian2016redirected,li2021openrdw}.

When comparing the performance of the RDW controllers introduced earlier, most of them use the increased movable distance through their RDW methods and the number of resets when a collision occurs~\citep{li2021openrdw,strauss2020steering}. These measurements are not surprising because the RDW controller's primary goal is to allow users to travel around virtual spaces larger than their tracked space~\citep{bozgeyikli2019locomotion,langbehn2017application}. However, another potential of RDW is that the semantic information between real and virtual spaces can also be matched when the coordinate systems of the two spaces are aligned, which means that physical components in the real space can be utilized in VR. Furthermore, since RDW using translation gain only controls perceived speed~\citep{interrante2007seven,selzer2022analysis,kim2021adjusting}, it can be used to rescale the virtual space while maintaining the coordinate system between the physical and virtual space. Therefore, this study focused on the adaptive application of RDW that can match the virtual space with the user's actual space.

\subsection{Cognitive Threshold and Visual Perception}

As RDW visually adjusts VR users' locomotion in VR, the redirecting gain value should be varied within the threshold range where they do not notice a significant change in their moving speed. Since Steinicke et al.~\citep{steinicke2009estimation} first estimated the RDW threshold, several studies have revisited the rotational, translational, and curvature gain threshold range~\citep{grechkin2016revisiting,zhang2018detection,rietzler2018rethinking}. Previous studies have shown that users tend to underestimate their speed when walking in a virtual environment while wearing VR HMDs~\citep{ash2013vection,banton2005perception,bruder2011tuning,caramenti2018matching}. Therefore, it is possible to transform users to move faster and wander around a more expansive space. Although RDW techniques utilize the user's visual perception in VR, various factors affect on RDW threshold range, such as the visual compositions of VEs and the sense of embodiment~\citep{nguyen2020effect,kruse2018can,neth2012velocity,paludan2016disguising}. In addition, the cognitive threshold range is also affected by the hardware capability of VR HMD, such as the Field of View (FoV)~\citep{williams2019estimation}.

In particular, the RDW threshold value is affected by parameters related to cognitive ability in VR. When applying RDW to the user, gradually increasing the RDW gain value can change their locomotion more than the initially recognized range~\citep{neth2012velocity,zhang2014human,sakono2021redirected}. Moreover, it is influenced by events occurring in virtual reality space. For example, when performing other tasks while walking, the user's attention was distributed to the task performance, making them less sensitive to changes in their speed~\citep{ciumedean2020mission,gao2020visual}. A study also shows that events in VR can act as distractors, dispersing users' attention and enlarging the transformable range~\citep{cools2019investigating,williams2019estimation,kim2022effects}. In order to adaptively transform space into a virtual environment, research on the effect of the composition of virtual space on RDW is required. Nguyen et al.~\citep{nguyen2018effect}'s study shows that the size of the virtual environment did not affect the threshold of curvature gain. In contrast, Kim et al.~\citep{kim2021adjusting} show that the translation gain threshold range increased in the larger virtual space. Even though the purpose of RDW is to match the tracked space to the virtual target space, studies focused on the effects of the spatial configuration of virtual space on the RDW threshold range were lacking.

\subsection{Relative Translation Gains}

We proposed Relative Translation Gains (RTGs) to adaptively transform VR user's space by adjusting the user's speed in virtual space by applying different translation gain values to width and depth, respectively~\citep{kim2021adjusting}. Unlike other RDW controllers that continuously change the gain values to redirect a user, RTG is applied only once according to the target space, which has less impact on possible VR simulator sickness than other RDW methods. The threshold range of RTG is defined through $\alpha_T$, which is the ratio of the two translation gains in a single pair of relative translation gains and is expressed by the following formula:

\begin{equation}
    \ \alpha_T = \frac{g_{T,x}}{g_{T,y}}\
    \label{equ:relativetranslation}
\end{equation}

where $g_{T,x}$ is the VR environment's x-axis translation gain, and $g_{T,y}$ is the VR environment's y-axis translation gain used as the reference translation gain. The RTG threshold range is defined as the value between the maximum 2D translation ratio and the minimum 2D translation ratio. By utilizing two translation ratio values and the translation gain threshold range from Steinicke et al.~\citep{steinicke2009estimation} 's study together, the moving speed can be adjusted at different ratios for the two axes in virtual reality. Through RTG application, the shape of the virtual space can be deformed more adaptively according to the target virtual space's configuration.

As RTG was proposed to transform movable space by controlling the user's speed in VR, studies should be conducted on how virtual space components affect the thresholds. When we first proposed RTG and estimated the threshold range, we experimentally revealed that the threshold range increased in a large-sized virtual room than in the smaller one~\citep{kim2021adjusting}. Furthermore, we discussed the different locations of the virtual horizon according to the size of the virtual room affecting the subjects' visual perception of their walking speed~\citep{messing2005distance}. Based on previous findings, we wanted to reveal how the subject's gaze distribution varies depending on the composition of the virtual space and how it relates to the RTG threshold. In this study, we focused on integrating fragmented implications about the effects of virtual space on the RTG threshold range to structurally explore the effect of virtual space configuration on RTG thresholds and propose a method of adapting RTG for space modification.

\section{Methodology}
\label{sec:3}

\subsection{Research Questions and Hypothesis}

This study aims to investigate the effect of the virtual space's configuration on the user's cognitive threshold when adjusting their speed using Relative Translation Gains (RTGs) and derives guidelines on utilizing RTGs as a spatial deformation component. We examined three factors of spatial configuration: the size of the space, the existence of objects, and spatial layout. Based on previous studies~\citep{kim2021adjusting,kim2022effects} that found user's visual perception changes according to the size and presence or absence of objects that affect the RTG threshold range and gaze distribution, we set our research questions as follows:

\,

\begin{enumerate}  [label={RQ\arabic*}.,noitemsep]
\item How do configurations of the virtual space affect the threshold range of relative translation gains?
\item How do configurations of the virtual space affect the subjects' gaze distribution?
\end{enumerate}

\,

To answer these research questions, we conducted two studies to derive structured guidelines for adaptively setting the RTG threshold range according to the spatial configuration. We utilized our previous study results (Study 1)~\citep{kim2022effects} and conducted a new study focused on spatial configuration (Study 2) to derive more structured guidelines for adaptively setting the RTG threshold range according to the spatial configuration. Study 1 was conducted for six conditions according to three room sizes (Large, Medium, Small) and the existence or absence of objects (Empty, Furnished). Since six conditions in Study 1 eventually changes the visibility level of space, we selected spatial layout as a variable to find out more about the impact of visibility on RTG threshold. Study 2 was conducted for eight conditions according to two room sizes (Large, Small) and four spatial layouts (Empty, Centered, Scattered, Peripheral). Through Study 2, we examined the effect of various spatial layouts configured by the arrangement of objects on the RTG threshold in a realistic scenario. Two size conditions from Study 1 (Small, Large) were used in Study 2 to compare the results between the two studies. In the case of spatial layout, we chose four spatial layouts to change the visibility level of each virtual space.

Based on our previous results that participants were less sensitive to being faster in a larger virtual room~\citep{kim2021adjusting}, we could set our first hypothesis about the room size. Moreover, we referred to Shin et al.~\citep{shin2021user}'s study to distinguish the level of visibility between four layout conditions to set our hypothesis about spatial layout. We assumed that the spatial layout with low visibility might distract users' attention more and make users less sensitive to being faster. For the last hypothesis about AoD, we assumed that the AoD distribution in the smaller room is lower than in the large one because the smaller room has a lower virtual horizon~\citep{messing2005distance}. Finally, we could complete the following three hypotheses:

\,
\begin{enumerate} [label={H\arabic*}.,noitemsep]
	\item The probabilities of ``faster'' responses to all relative translation gains will be lower in the large size conditions than in smaller conditions.
	\item The probabilities of ``faster'' responses to all relative translation gains will be lower in the spatial layouts with low visibility.
    \item The medians of AoD from participants will be lower in the large size conditions than in smaller conditions.
\end{enumerate}

\,

\subsection{Study Design}

\subsubsection{Study 1: Size $\times$ Object Existence}

\begin{figure}[tb]
 \centering
 \includegraphics[width=\columnwidth]{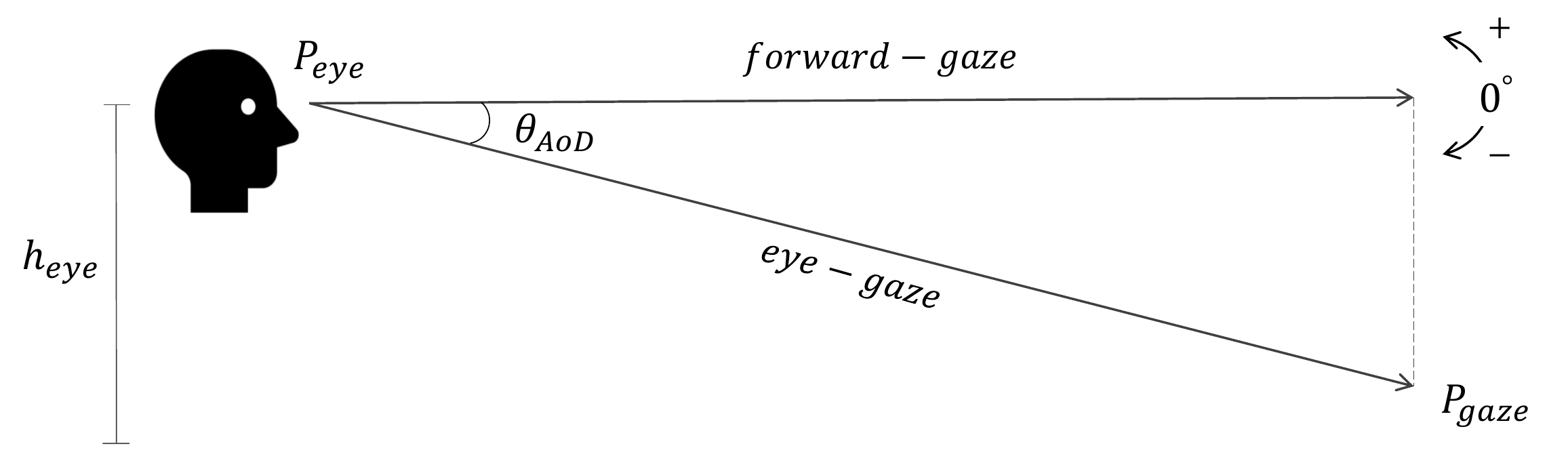}
 \caption{The Angle of Declination (AoD) between the participant's eye-gaze and forward-gaze. Forward-gaze refers to the orthogonal projection vector of the eye-gaze.}
 \label{fig:AoD}
\end{figure}

\begin{figure}[t]
 \centering
 \includegraphics[width=\columnwidth]{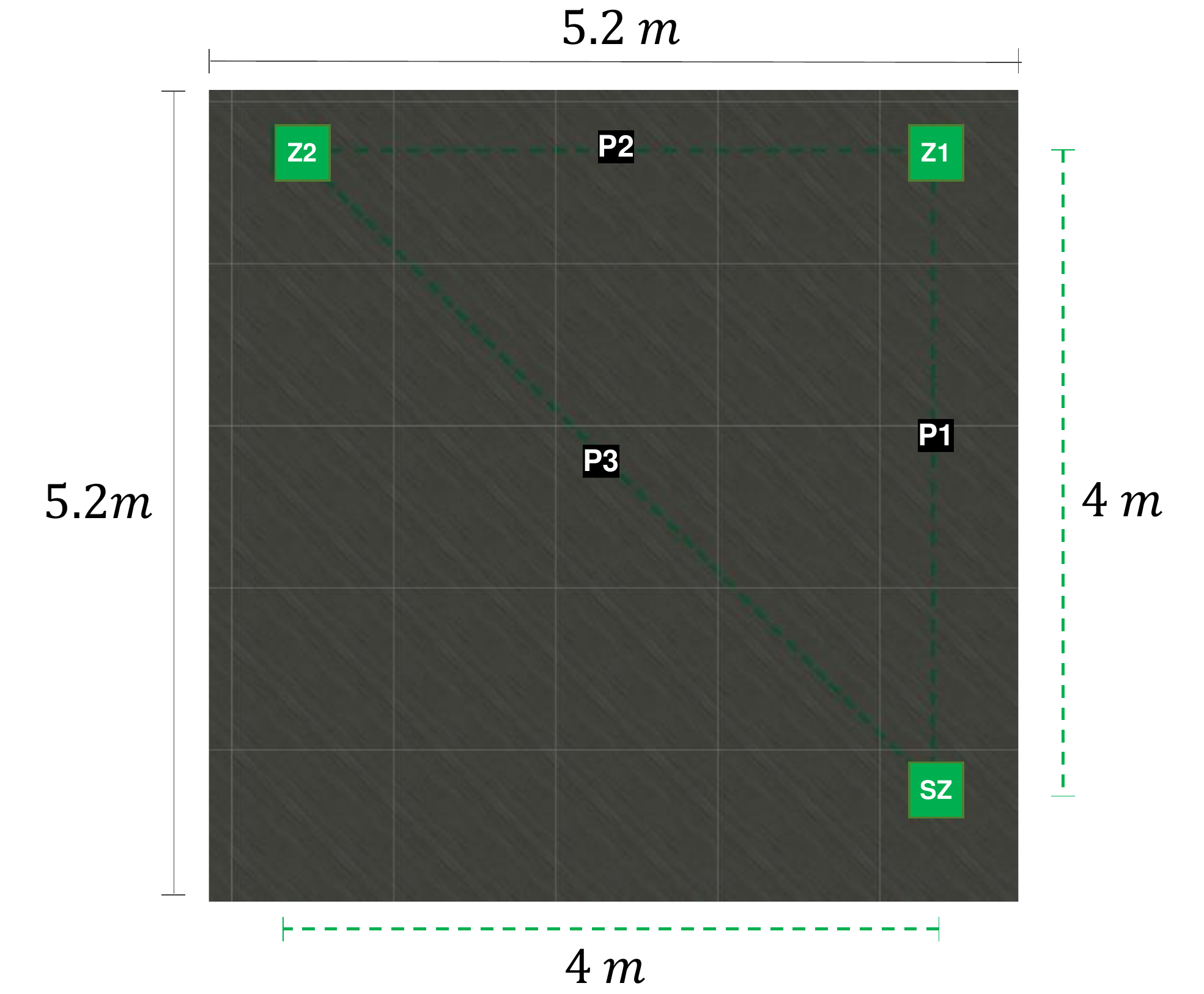}
 \caption{Top-view of the path in Small $\times$ Empty condition in Study 2. Participants walk along the path repeatedly in order of P1, P2, and P3. (SZ = Start Zone, Z1 = Zone 1, Z2 = Zone 2).}
 \label{fig:pathDesign}
\end{figure}

\begin{figure*}[t]
 \centering
 \includegraphics[width=\textwidth]{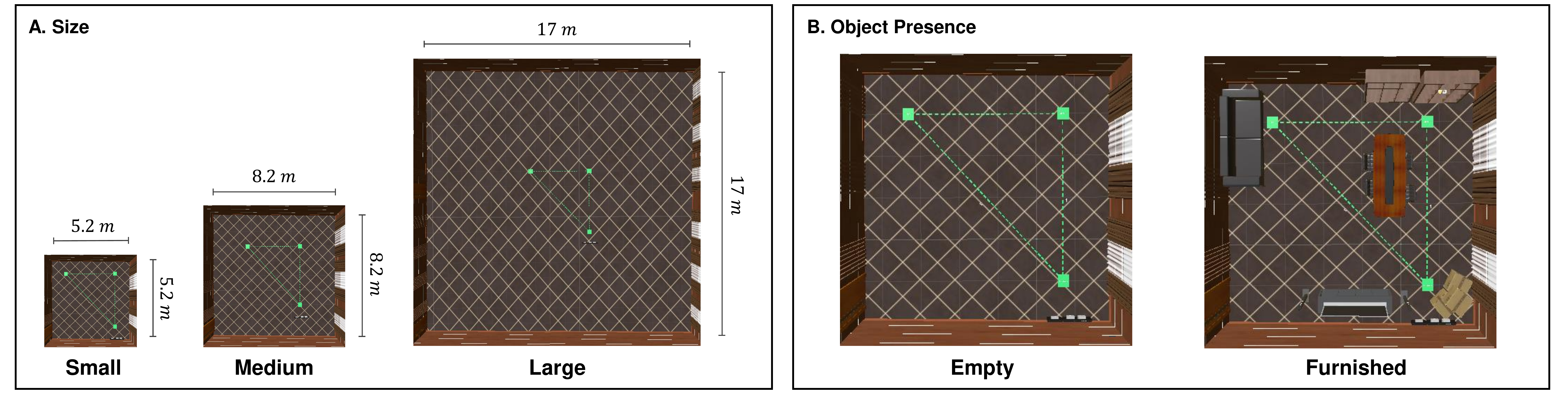}
 \caption{(A) Three size conditions used in Study 1. (B) Top-view of the Empty and the Furnished condition in the Small size condition.}
 \label{fig:studydesign1}
\end{figure*}

\begin{figure*}[t]
 \centering
 \includegraphics[width=\textwidth]{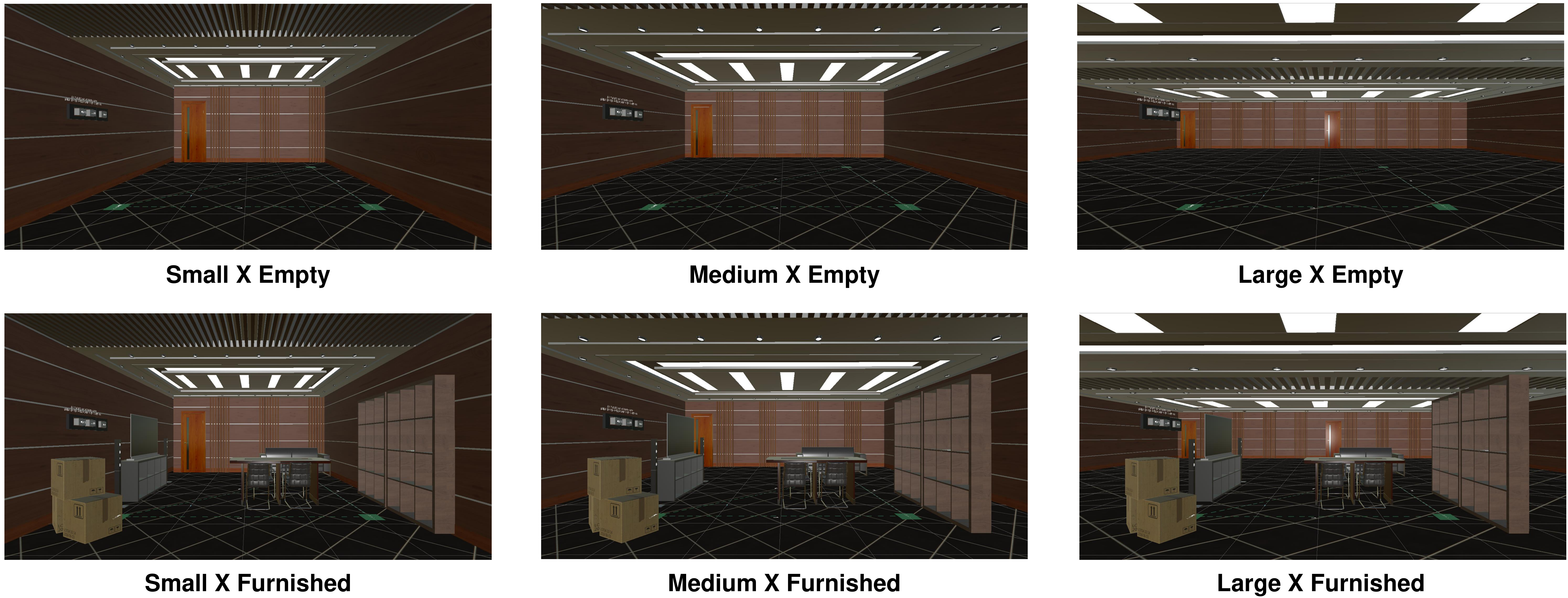}
 \caption{Screenshot of the six conditions in Study 1. Combined path and furniture are placed at the center of each condition.}
 \label{fig:conditions1}
\end{figure*}

We used three out of five paths used in a previous study~\citep{kim2021adjusting}, taking into account the location of objects in the Furnished condition. Fig.~\ref{fig:pathDesign} shows the three pathways we used in our study. Subjects repeatedly performed the walking task in sequence from Path 1 to Path 3. Through Path 1 and Path 2, participants experienced maximum and minimum translation gains. In Path 3, they experienced the square mean of the maximum and minimum translation gains. We fixed the reference translation gain ($g_{T,y}$) at 1.0. To estimate 2D translation ratio, another axis’s translation gain ($g_{T,x}$) was set at 0.825, 0.875, or 0.925 for minimum ratio, and 1.15, 1.2, or 1.25 for maximum ratio. These six relative translation gains were randomly ordered and repeated seven times with the Latin-squared method for within-counterbalancing.

As shown in Fig.~\ref{fig:AoD}, the AoD ($\theta_{AoD}$) between eye-gaze and the forward-gaze can be measured from the position of the eye-gazed point ($P_{gaze}$) and the position of the eye ($P_{eye}$). We can calculate $\theta_{AoD}$ with the following formula:
\begin{equation}
    \ \theta_{AoD} = \arctan (\frac{y_{gaze} - y_{eye}}{\sqrt{(x_{gaze} - x_{eye})^2 + (z_{gaze} - z_{eye})^2}})
    \label{equ:theta_AoD}
\end{equation}

where the position of the eye-gazed point is $P_{gaze}$ = ($x_{gaze}$, $y_{gaze}$, $z_{gaze}$) and the position of the eye is $P_{eye}$ = ($x_{eye}$, $y_{eye}$, $z_{eye}$). Three size conditions were generated to differing average AoD values, and the location of objects was also set to differing the AoD distribution~\citep{kim2022effects}. Fig.~\ref{fig:studydesign1} shows three size conditions and two object layouts used in Study 1. Three room sizes (Large, Medium, Small) and two object layouts (Empty, Furnished) were combined to generate six experimental conditions for estimating RTG thresholds, as shown in Fig.~\ref{fig:conditions1}. Moreover, we added two conditions without RTG to determine the effect of RTG on VR Simulator Sickness Questionnaire (SSQ) scores~\citep{kennedy1993simulator}. Our previous study~\citep{kim2021adjusting} found that post-SSQ was higher in a larger virtual space, so we chose the Large condition to observe the effect of RTG on the VR SSQ score. We generated eight experimental conditions as follows: Small $\times$ Empty, Medium $\times$ Empty, Large $\times$ Empty, Large $\times$ Empty (No RTGs), Small $\times$ Furnished, Medium $\times$ Furnished, Large $\times$ Furnished, Large $\times$ Furnished (No RTGs).

\subsubsection{Study 2: Size $\times$ Spatial Layout}
\label{sec:3.2.2}

\begin{figure*}[t]
 \centering
 \includegraphics[width=\textwidth]{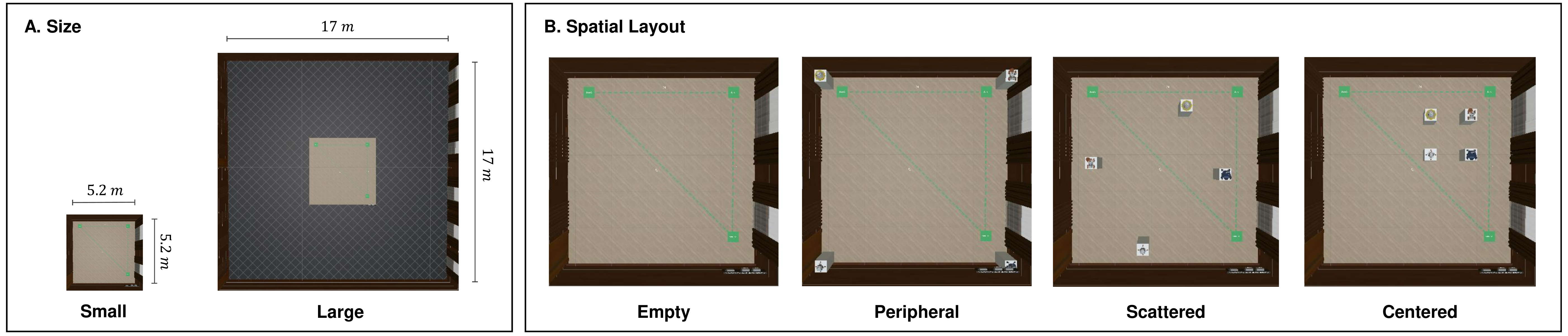}
 \caption{(A) Two size conditions used in Study 2. (B) Top-view of the four spatial layout conditions in the Small size condition.}
 \label{fig:studydesign2}
\end{figure*}

\begin{figure*}[t]
 \centering
 \includegraphics[width=\textwidth]{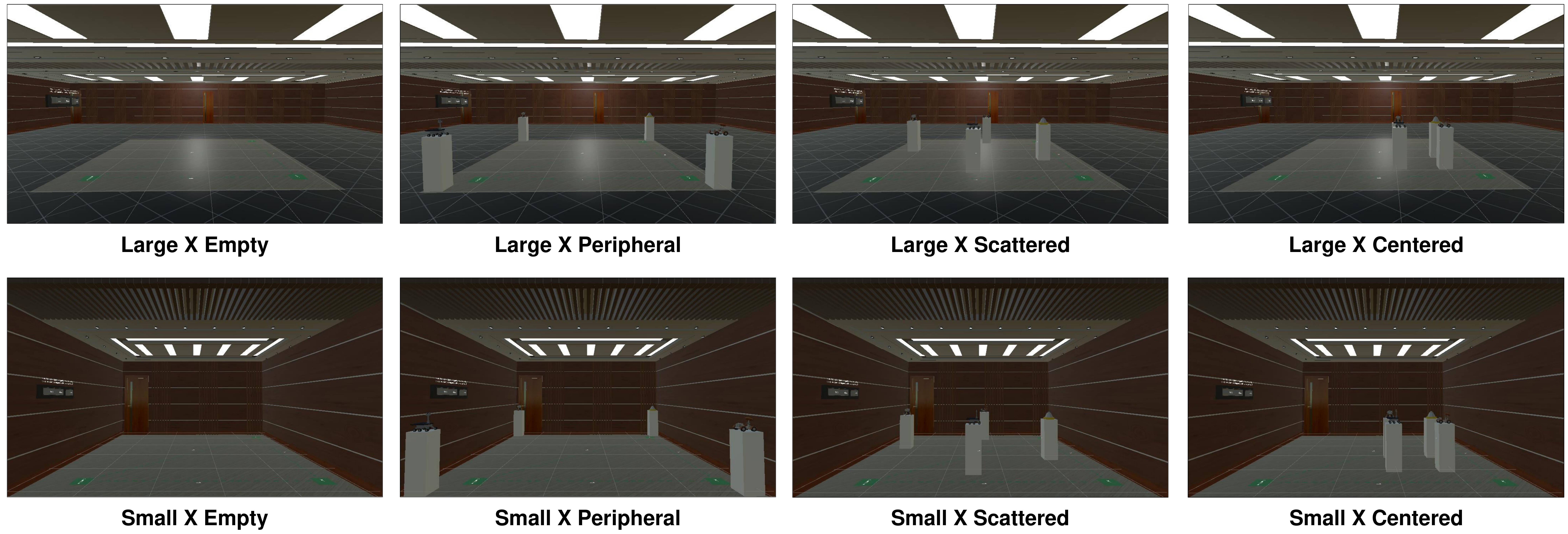}
 \caption{Screenshot of the eight experimental conditions in Study 2. The inner floor area expressed in the Large condition is the same size as the Small conditions' entire floor area and is located at the center of the room. Combined paths and exhibit objects are placed in it.}
 \label{fig:conditions2}
\end{figure*}

In Study 2, we examined the effect of spatial layout according to object placement on the RTG threshold range. The experiment employed a mixed-subject method with the size of the virtual room as the between-subject variable and the spatial layout of the VR room as the within-subject variable. In designing the experiment conditions for Study 2, we decided to use two size conditions (Small, Large) from Study 1 as shown in Fig.~\ref{fig:studydesign2} (A). To analyze the effect of spatial layout on RTG thresholds, we decided to reduce the number of objects compared to Study 1. This is because when objects are placed densely, it is highly likely that they may distract users' attention regardless of the spatial layout they configured. To construct a realistic virtual space that is relatively less crowded, we assumed an exhibition room set-up consisting of four exhibition stands with 3D spacecraft models with similar volumes placed on them. This was to minimize the difference between each set of objects, whose locations determined the spatial layout so that the level of distraction caused by their existence could be controlled.

In setting the spatial conditions, we visualized the adjusted movable space by applying the concepts proposed in Study 1~\citep{kim2022effects}. The inner floor area in the Large condition, which was the same size as the entire floor area of the Small condition, was expressed with different-colored tiles to show where subjects could walk, as shown in Fig.~\ref{fig:conditions2} (A). In the Small condition, the perceived movable space was the same as the adjusted movable space. Based on the Empty layout, four exhibition stands were placed in different locations to create three spatial layouts, following Shin et al.~\citep{shin2021user}'s study to set different levels of visibility afforded in each space. We selected Peripheral, Scattered, and Centered layouts for corresponding visibility levels: Low, Medium, and High. In the Peripheral condition, exhibition stands were placed at the four corners of the adjusted movable space (Fig.~\ref{fig:studydesign2} (B) Peripheral). In the Scattered conditions, two exhibition stands were placed inside the walking path, which was identical to Study 1, and the other two exhibition stands were spread outside the path (Fig.~\ref{fig:studydesign2} (B) Scattered). Lastly, all four exhibition stands were gathered inside the triangle created by the walking path in the Centered layout (Fig.~\ref{fig:studydesign2} (B) Centered). By combining two size conditions (Small, Large) and four spatial layouts (Empty, Peripheral, Scattered, Centered), we generated eight experimental conditions to estimate the RTG threshold range, as illustrated in Fig.~\ref{fig:conditions2}.

\subsection{System Implementation and Setup}

Both studies were set up in an empty 6 m $\times$ 6 m physical indoor space with four HTC VIVE base stations (v2.0) installed at the top corners. The subjects wore an HTC VIVE Pro Eye with HTC VIVE wireless adaptor. They held the VIVE controller in their hands and used the controller to press a button in the virtual scene. The virtual experience was operated using a desktop computer consisting of an i9-10900K 3.70GHz CPU, 128GB RAM, and a GeForce RTX 3090 24GB GPU. To implement a virtual conference room environment, we developed our system with Unity 3D (v2019.3.7f1) and steam VR Plugin (v1.16.10). Every virtual room condition was created from the same conference room prefab. We employed SRanipal SDK and TobiiXR Plugin to get each subject's eye-tracking data. We obtained the subject's AoD through Equation~\ref{equ:theta_AoD}, which was approximately 36 $\theta_{AoD}$ samples per second. From the accumulated AoD values, we extracted AoD data obtained from each zone where users made a change in their paths or stood at a fixed position to answer questionnaires. In Study 2, the location gazed by the subjects on the entire virtual room was acquired to draw the subjects' gaze heatmap. In order to understand the areas where the subjects frequently gazed, we measured the time the users' gazes hit the four walls, floors, and objects, respectively. In particular, we measured the time users gaze at the inner and outer floor areas separately in the Large size conditions to focus more on the inner floor area where objects were placed. We then used Heatmapper~\citep{babicki2016heatmapper} to draw gaze heatmaps from the measured subjects' gaze data.

\subsection{Participants}

\begin{table}
  \caption{Study 1 participants' information in two groups}
  \label{tab:Study1participant}
  \scriptsize%
	\centering%
  \begin{tabu}{%
	*{7}{c}%
	*{2}{r}%
	}
  \toprule
   Participants & Group1 &  Group2 \\
  \midrule
  Between Condition & Empty & Furnished \\
  Number of Male & 10 & 10  \\
  Number of Female & 6 & 6  \\
  Avg. Height (Male) & \begin{tabular}[c]{@{}c@{}}175.4 cm\\ (SD = 5.2)\end{tabular} & \begin{tabular}[c]{@{}c@{}}175.1 cm\\ (SD = 7.4)\end{tabular} \\
  Avg. Height (Female) & \begin{tabular}[c]{@{}c@{}}163.2 cm\\ (SD = 6.5)\end{tabular} & \begin{tabular}[c]{@{}c@{}}163.3 cm\\ (SD = 1.9)\end{tabular}\\
  \midrule
  \end{tabu}%
\end{table}

In Study 1, we recruited 32 subjects through the local university website and paid \$ 20 in remuneration. Twenty of them identified as male and 12 as female. All subjects were at least 18 years old, and all had normal vision or corrected to normal vision. The subjects' mean age was 23.81 (SD = 3.86), and their mean interpupillary distance (IPD) of them was 64.17 (SD = 1.89). Most of them had a moderate level of experience in HMD-mediated VR environments: 24 participants had worn VR HMDs up to ten times before and four for more. Only four of them had no prior VR experience. We separated subjects into two groups to decrease fatigue and learning during experiments with VR HMDs. We maintained the gender ratio among the groups that might affect the cognitive threshold of redirected walking~\citep{williams2019estimation}. Furthermore, as we assumed the AoD distribution might affect the threshold range, we balanced the average height between the two groups. The detailed information of the two groups that participated in Study 1 is shown in Table~\ref{tab:Study1participant}.

\begin{table}
  \caption{Study 2 participants' information in two groups}
  \label{tab:study2participant}
  \scriptsize%
	\centering%
  \begin{tabu}{%
	*{7}{c}%
	*{2}{r}%
	}
  \toprule
   Participants & Group1 &  Group2 \\
  \midrule
  Between Condition & Large & Small \\
  Number of Male & 8 & 8  \\
  Number of Female & 8 & 8  \\
  Avg. Height (Male) & \begin{tabular}[c]{@{}c@{}}173.5 cm\\ (SD = 6.9)\end{tabular} & \begin{tabular}[c]{@{}c@{}}173.6 cm\\ (SD = 6.4)\end{tabular}\\
  Avg. Height (Female) & \begin{tabular}[c]{@{}c@{}}163.1 cm\\ (SD = 5.9)\end{tabular} & \begin{tabular}[c]{@{}c@{}}163.6 cm\\ (SD = 5.6)\end{tabular} \\
  \midrule
  \end{tabu}%
\end{table}

In Study 2, the subjects were recruited through the local university website and paid \$20 in remuneration. Of the 32 subjects, 16 identified as male and 16 as female. All subjects were at least 18 years old, and all had normal vision or corrected normal vision. The mean age of the subjects was 22.97 (SD = 3.32), and their mean interpupillary distance (IPD) of them was 64.01 (SD = 1.79). Most subjects had a moderate level of experience in HMD-mediated VR environments: 23 subjects had worn VR HMDs up to ten times before and four for more. Only five of them had no prior experience. As in Study 1, we separated subjects into two groups and balanced the average height and gender ratio between the two groups, as shown in Table~\ref{tab:study2participant}.

\subsection{Tasks and Procedures}

An Institutional Review Board approved the content and procedures of both Study 1 and Study 2. Participants first wore an HTC VIVE Pro Eye and held the VIVE controller to complete an eye-tracking calibration session. They then conducted test trials before the main study trials, which included walking along the three given paths in sequential order. After they walked along the three paths, they were asked to answer whether their movements in the virtual environment were ``faster'' or ``slower'' than in the real environment. We used a pseudo-Two-Alternative Forced-Choice (pseudo-2AFC), which is widely used for estimating redirected walking threshold range~\citep{grechkin2016revisiting, steinicke2009estimation, williams2019estimation}. After the test trial, they took off the VR HMD and filled out the pre-SSQ. Next, participants conducted the main trials. They walk along the sequential path while experiencing changes to their perceived speed between the maximum and minimum translation gain values. They repeated 42 trials (six relative translation gains $\times$ seven repetitions) by pressing the ``next'' button. They were permitted to take breaks whenever they wanted.

We created three different sizes of virtual rooms, Large (17 m $\times$ 17 m), Medium (8.2 m $\times$ 8.2 m), and Small (5.2 m $\times$ 5.2 m), as shown in the Fig.~\ref{fig:studydesign1}. We set up six conditions by combining three size conditions (Large, Medium, Small) with the presence of objects (Furnished, Empty) as shown in Fig.~\ref{fig:conditions1}. The four conditions conducted in each participant group were randomized and counterbalanced. Participants were required to answer the post-SSQ after each condition was completed. After they finished every condition, they were subject to a short semi-structured interview on how the VR room’s configuration affected their response to the questionnaire. Each group of participants experienced 168 trials (42 trials $\times$ four experimental conditions) and took approximately 15 minutes for each experimental condition. The total duration of each study session was approximately two hours.

The overall tasks and procedures of Study 2 were the same as Study 1, except for the experimented conditions in each group. Subject group 1 experienced four spatial layout conditions (Empty, Centered, Peripheral, Scattered) in a Large (17 m $\times$ 17 m) virtual room, and the other group experienced four spatial layout conditions (Empty, Centered, Peripheral, Scattered) from Small (5.2 m $\times$ 5.2 m) virtual room.

\section{Results}
\label{sec:4}

\subsection{Statistic Results}

\begin{figure*}[tb]
 \centering
 \includegraphics[width=\textwidth]{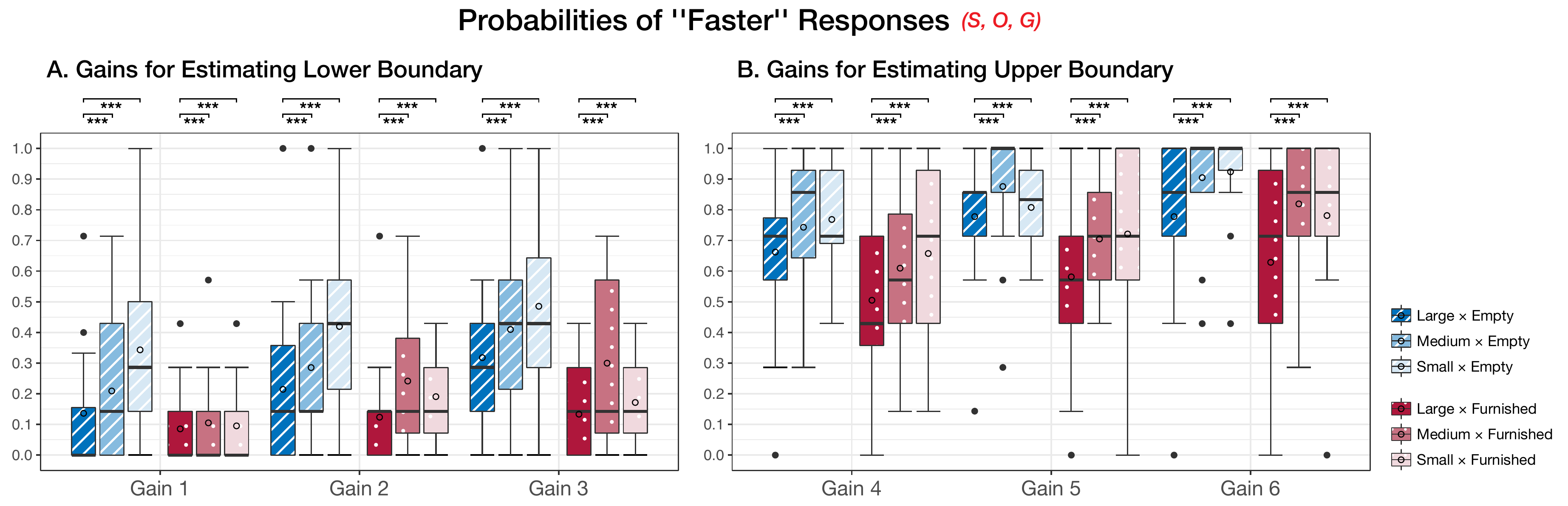}
 \caption{The effects of Size and Object Existence on the probabilities of ``faster'' responses (o: mean)}
 \label{fig:ARTResult1}
\end{figure*}

We evaluated how the VR room size (\textit{Size}) and object presence (\textit{Object}) affect the user's relative translation gain thresholds (\textit{Gain}) by comparing the probabilities of ``faster'' responses from our previous Study 1~\citep{kim2022effects} (\textit{Size}: $F(2,490) =$ 16.690, $p<$ .001; \textit{Object}: $F(1,490) =$ 60.984, $p<$ .001). As shown in Fig.~\ref{fig:ARTResult1}, the mean value for the probability of ``faster'' responses was significantly lower in the Large room than both the Medium and Small room for all six relative translation gains (Large-Medium: $p<$ .001; Large-Small: $p<$ .001). Moreover, the mean probability value for the Furnished room was also significantly lower than for the Empty room across all gains ($p<$ .001).


\begin{figure*}[tb]
 \centering
 \includegraphics[width=\textwidth]{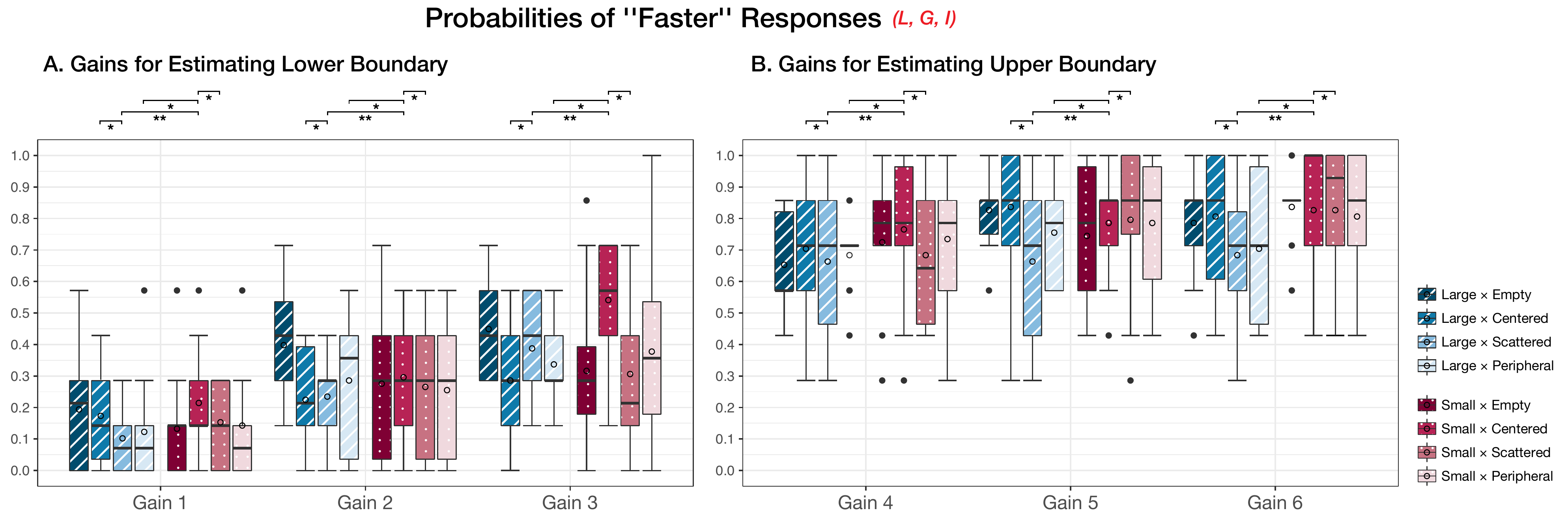}
 \caption{The effects of Size and Spatial Layout on the probabilities of ``faster'' responses (o: mean)}
 \label{fig:ARTResult2}
\end{figure*}

We investigated how the VR room size (\textit{Size}) and spatial Layout (\textit{Layout}) affect the user's relative translation gain thresholds (\textit{Gain}) by comparing the probabilities of ``faster'' responses in Study 2. Same as Study 1, the ART for non-parametric factorial ANOVA analysis ($\alpha = 0.05$) was adapted to conduct the multivariate analysis within the within-subject factor of \textit{Layout}, the between-subject factor of \textit{Size}, and the factor of \textit{Gain}. The ART-C algorithm~\citep{elkin2021aligned} was also used for the multifactor post hoc contrast tests, and all pairwise comparisons were Bonferroni corrected. \textit{Size} consists of Large and Small, and \textit{Layout} consists of Empty, Centered, Peripheral, and Scattered conditions described in Section~\ref{sec:3.2.2}. Six relative translation gain factors represent the following pairs of x-axis and y-axis translation gain value: Gain 1 = (0.825, 1), Gain 2 = (0.875, 1), Gain 3 = (0.925, 1), Gain 4 = (1.15, 1), Gain 5 = (1.20, 1), and Gain 6 = (1.25, 1). We excluded four participants' data as outliers; The two in the Small room condition and the other two in the Large room condition were extreme values or considered contaminated, recorded as zero.

We found significant main effects of \textit{Layout} ($F(3,611) =$ 3.198, $p=$ .023) and \textit{Gain} ($F(5,611) =$ 228.642, $p<$ .001). However, there was no significant main effects of \textit{Size} ($F(1,611) =$ 3.157, $p=$ .076) was founded. The pairwise comparison revealed significant differences between pairs of \textit{Size} condition: Centered and Scattered ($p=$ .035). However, a significant difference was not shown between the other two pairs of spatial layouts (all $p>$ .05). We also found a significant interaction of \textit{Size} and \textit{Layout} ($F(3,611) =$ 2.941, $p=$ .033). The post hoc analysis revealed the significant differences between Large $\times$ Scattered and Small $\times$ Centered ($p=$ .003) and between Large $\times$ Peripheral and Small $\times$ Centered ($p=$ .035). However, it was not found any significant differences in the pairs of \textit{Size} $\times$ \textit{Layout}, nor significant main interaction effects between other factors (all $p>$ .05). Therefore, as shown in Fig.~\ref{fig:ARTResult2}, we found that the mean value for the probability of ``faster'' responses was significantly lower in the Scattered Layout than in the Centered Layout for all six relative translation gains. The mean probability value in the Large $\times$ Scattered room was significantly lower than the Small $\times$ Centered room for all gains. Moreover, the mean probability value in the Large $\times$ Peripheral room was significantly lower than the Small $\times$ Centered room for all gains.

\subsection{Threshold Estimation}

\begin{figure*}[tb]
 \centering
 \includegraphics[width=\textwidth]{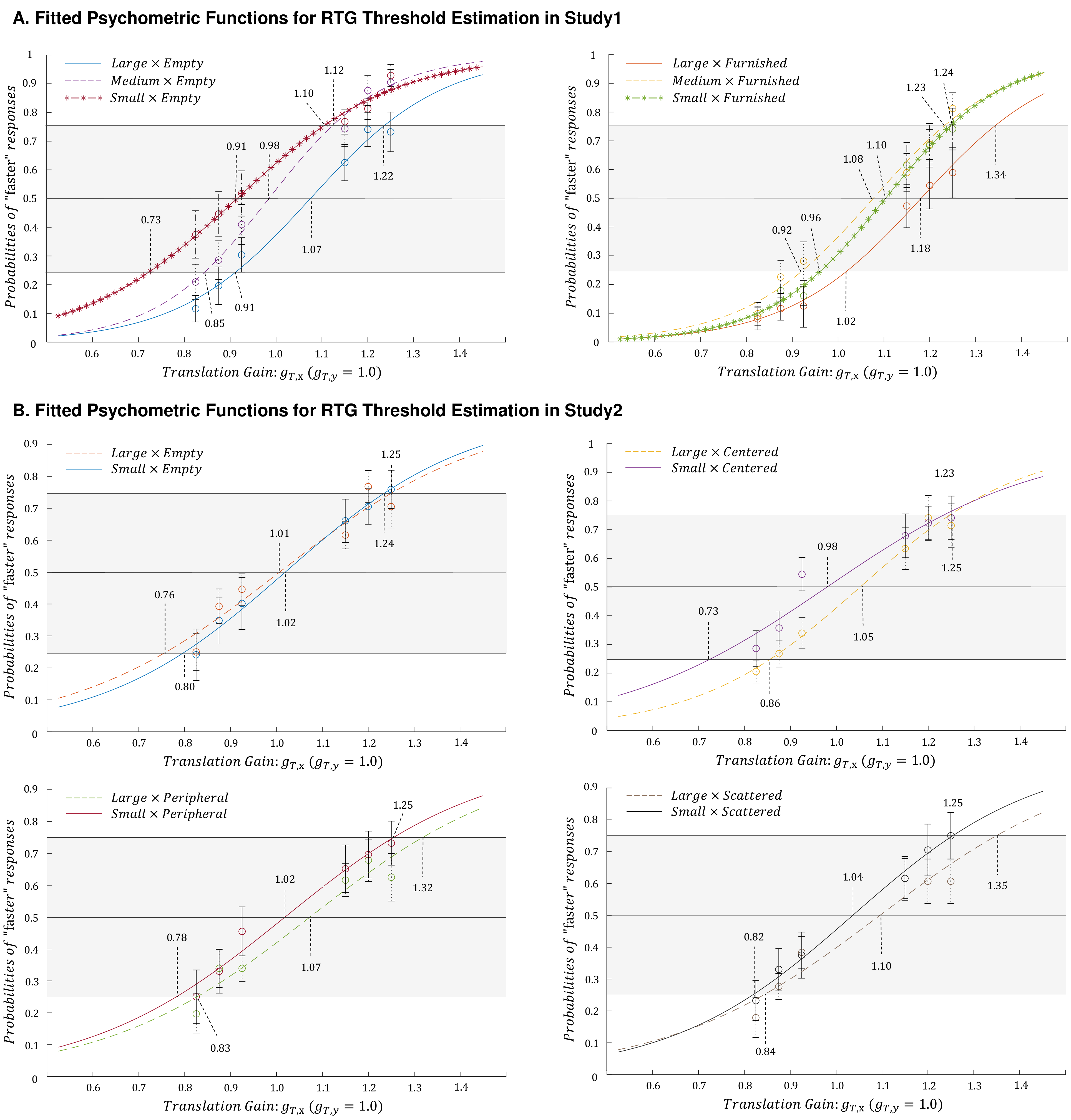}
 \caption{Fitted psychometric functions of mean estimated threshold values in six conditions from (A) Study 1 and (B) Study 2.}
 \label{fig:threshold}
\end{figure*}

We plotted fitted psychometric functions of mean estimated threshold values from Study 1 and Study 2. Fig.~\ref{fig:threshold} shows the mean estimated threshold values and Standard Error of the Mean (SEM) for the experimented relative translation gains. The graph's x-axis shows the translation gain for the x-axis ($g_{T,x}$) of the virtual room where the reference translation gain ($g_{T,y}$) is fixed to 1.0. The graph's y-axis shows the probabilities of ``faster'' responses to the questionnaire, ``Was the virtual movement faster or slower than the physical movement?''. We used a standard logistic psychometric function to fit the data, and it can be represented as follows:

\begin{equation}
    \ f(x)=\frac{1}{1+e^{ax+b}}\
    \label{equ:pse}
\end{equation}

The Point of Subjective Equality (PSE), the value at which a user perceives the virtual speed to be identical to their actual speed, is obtained from the fitted psychometric functions. We used the 25\% and 75\% criterion used from Steinicke et al.~\citep{steinicke2009estimation} to estimate the RTG's threshold value (2D translation ratio, $\alpha_T$). $\alpha_{T,lower}$ (minimum 2D translation ratio) refers to a lower boundary of the threshold range, and $\alpha_{T,upper}$ (maximum 2D translation ratio) refers to an upper boundary of the threshold range. When the participant's perceived speed of one axis in the virtual environment is adjusted to be slower or faster between $\alpha_{T,lower}$ and $\alpha_{T,upper}$, they rarely notice their speed differences.

\begin{table}[t]
  \caption{Relative translation gain thresholds according to virtual room configurations in Study 1.}
  \label{tab:threshold1}
  \scriptsize%
	\centering%
  \begin{tabu}{%
	*{7}{c}%
	*{2}{r}%
	}
  \toprule
   VE Configuration & $\alpha_{T,lower}$ &   PSE &   $\alpha_{T,upper}$ \\
  \midrule
  Large $\times$ Empty & 0.91 & 1.07  & 1.22 \\
  Medium $\times$ Empty & 0.85 & 0.98  & 1.12 \\
  Small $\times$ Empty & 0.73 & 0.91  & 1.10 \\
  Large $\times$ Furnished & 1.02 & 1.18  & 1.34 \\
  Medium $\times$ Furnished & 0.92 & 1.08  & 1.23 \\
  Small $\times$ Furnished & 0.96 & 1.10  & 1.24 \\
  \midrule
  \end{tabu}%
\end{table}

\begin{table}[t]
  \caption{Relative translation gain thresholds according to virtual room configurations in Study 2.}
  \label{tab:threshold2}
  \scriptsize%
	\centering%
  \begin{tabu}{%
	*{7}{c}%
	*{2}{r}%
	}
  \toprule
   VE Configuration & $\alpha_{T,lower}$ &   PSE &   $\alpha_{T,upper}$ \\
  \midrule
  Large $\times$ Empty & 0.76 & 1.01  & 1.25 \\
  Large $\times$ Centered & 0.86 & 1.05  & 1.25 \\
  Large $\times$ Peripheral & 0.83 & 1.07  & 1.25 \\
  Large $\times$ Scattered & 0.84 & 1.10  & 1.35 \\
  Small $\times$ Empty & 0.80 & 1.02  & 1.25 \\
  Small $\times$ Centered & 0.73 & 0.98  & 1.23 \\
  Small $\times$ Peripheral & 0.78 & 1.02  & 1.25 \\
  Small $\times$ Scattered & 0.82 & 1.04  & 1.25 \\
  \midrule
  \end{tabu}%
\end{table}

The estimated RTG thresholds from six virtual room conditions in Study 1 were written in Table~\ref{tab:threshold1}. As shown in Table~\ref{tab:threshold1}, estimated RTG threshold values increase according to the size of the virtual room became larger. Furthermore, the RTG threshold values were higher in Furnished conditions than in the Empty ones. Similarly, the estimated RTG threshold values from eight virtual room conditions in Study 2 were shown in Table~\ref{tab:threshold2}. Results show the maximum 2D translation gain boundary value $\alpha_{T,upper}$ in Large $\times$ Scattered was higher than the Large conditions with other spatial layouts. Moreover, we could found the RTG threshold values from the Small $\times$ Centered layout were lower than the Small $\times$ Empty layout.

\subsection{Gaze Distribution}

To answer our second research question (RQ2: How do configurations of the virtual space affect the subjects' gaze distribution?), we utilized the AoD distribution graphs and gaze heatmap. Through the AoD distribution graph, we try to figure out the relationship between the location of the virtual horizon and the RTG threshold range. Moreover, we used a gaze heatmap in Study 2 to observe how the effect as a gaze distributor of objects changes according to the spatial layout.

\subsubsection{AoD Distribution}

\begin{figure*}[tb]
 \centering
 \includegraphics[width=\textwidth]{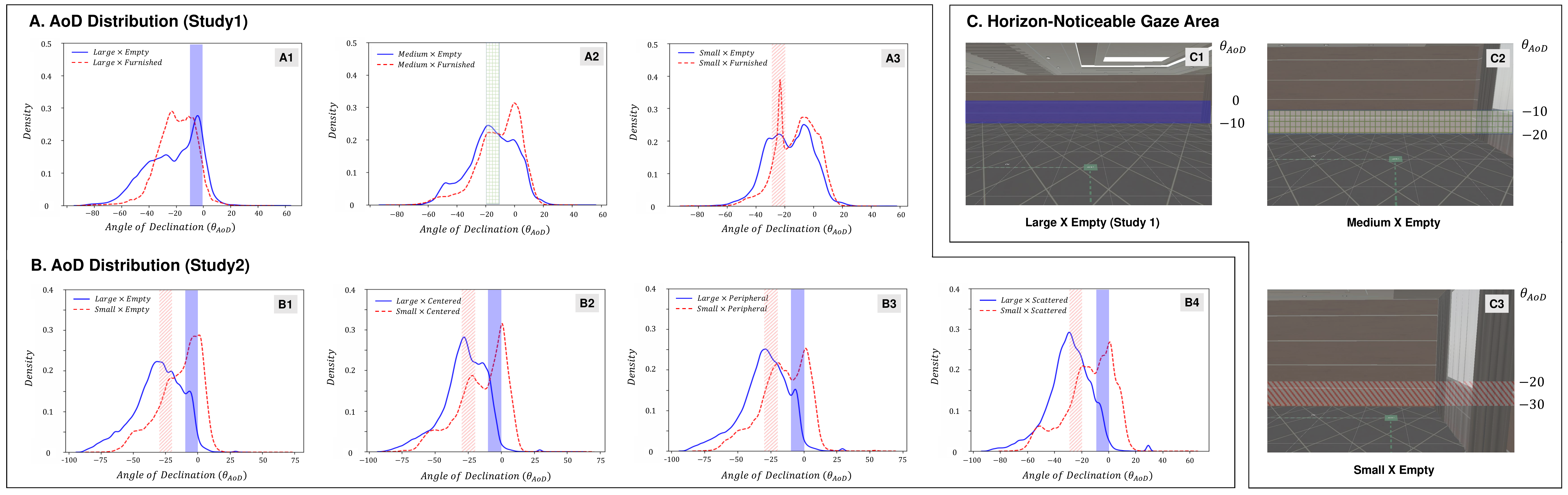}
 \caption{The density distribution of AoD according to VE configurations in (A) Study 1 and (B) Study 2. (C) The horizon-noticeable gaze area where users can perceive the virtual horizon for (C1) Large, (C2) Medium, and (C3) Small conditions. The highlighted area refers the horizon-noticeable gaze area in corresponding Empty conditions.}
 \label{fig:AoDResult}
\end{figure*}

Fig.~\ref{fig:AoDResult} shows the distribution of AoD according to each virtual room's conditions. Fig.~\ref{fig:AoDResult} (A) are the distribution of AoD from Study 1 and Fig.~\ref{fig:AoDResult} (B) are the distribution of AoD from Study 2. The x-axis of the AoD distribution graph is the AoD ($\theta_{AoD}$) between the user's eye-gaze and the forward-gaze. The graph's y-axis shows the density of normalized AoD for each subject. When a subject gaze straight forward while they are walking, $\theta_{AoD} = 0^{\circ}$. $\theta_{AoD}$ increases to a positive degree when they gaze upward, and on the contrary, $\theta_{AoD}$ decreases to a negative degree when they gaze downward. The AoD degree with high density means subjects gaze to that particular AoD degree frequently. As the AoD distribution graph aims to determine the relationship between the virtual horizon and the RTG threshold range, it is important to know how much the subjects perceived the virtual horizon in corresponding virtual room conditions. Humans typically recognize about 10 degrees downward from the location of the foveal area while they are walking~\citep{hill1986preferred}, so we set the horizon-noticeable gaze area of each room with this regulation. In the Small $\times$ Empty condition, the average AoD was set to $-20^{\circ}$ so we set the AoD area where the user could perceive the virtual horizon to be $-20^{\circ}$ to $-30^{\circ}$. Similarly, the horizon-noticeable gaze area in the Middle $\times$ Empty condition was $-10^{\circ}$ to $-20^{\circ}$, and in the Large $\times$ Empty conditions was $0^{\circ}$ to $-10^{\circ}$. In Fig.~\ref{fig:AoDResult}, we highlighted the AoD area of the Empty conditions where users could perceive the virtual horizon. The colored area in Fig.~\ref{fig:AoDResult} (C1) are the horizon-noticeable gaze area in Large $\times$ Empty, the dotted area in Fig.~\ref{fig:AoDResult} (C2) show those of Medium $\times$ Empty, and the dashed area in  Fig.~\ref{fig:AoDResult} (C3) show those of Small $\times$ Empty.

Fig.~\ref{fig:AoDResult} (A) shows that there is a high-density peak in the area where users can perceive the virtual horizon from each Empty condition from Study 1. In particular, this AoD peak is more pronounced in the Large sizes, as shown in Fig.~\ref{fig:AoDResult} (A1), consistent with the statistical results for AoD medians that the subjects look downward while they are walking in a larger room. Furthermore, Fig.~\ref{fig:AoDResult} (A1) shows that the existence of virtual objects biases the AoD distribution to be lower in Large conditions. In the Medium and Small virtual room, the Furnished condition also has a peak at higher AoD compared to the Empty one as shown in Fig.~\ref{fig:AoDResult} (A2),(A3). 

In order to more quantitatively analyze user gaze information, the Aligned Rank Transform (ART) for the non-parametric factorial ANOVA method ($\alpha = 0.05$) was adapted to handle the multivariate analysis, and all post hoc contrast comparisons were Bonferroni corrected. In Study 1, we investigated user's AoD values within the within-subject factor of \textit{Size} and the between-subject factor of \textit{Object}: The results only showed a significant main effect of \textit{Size} ($F(2,70) =$ 4.832, $p<$ .011). During the post hoc analysis, we found significant differences in pairs between Large and Medium ($p=$ .018), and Large and Small ($p=$ .043), but the pair of Medium and Small showed no significant difference ($p>$ .05). The other factor \textit{Object} nor main interaction showed no significant effects (\textit{Object}: $F(1,70) =$ 1.339, $p=$ .251; \textit{Size$\times$Object}: $F(2,70) =$ .172, $p=$ .842). The AoD distribution graph and statistical results confirmed that participants frequently notice the virtual horizon in Empty conditions, and they look more downward in a larger-sized condition than in a smaller one. On the other hand, the presence of objects distributes users' attention, which weakens the relationship between AoD distribution and the room size in the Furnished conditions.

The AoD distribution graph from Study 2 showed that the percentage of horizon-noticeable gaze area, which is determined by the size of the virtual space, has no distinguishable peak in corresponding size conditions. They were not much different from the density distribution of horizon-noticeable gaze area in the Large $\times$ Empty condition. These AoD results from Large conditions imply that an inner floor area with different tiles functions as a distractor, distributing the user's gaze to reach the virtual horizon configured by the wall and the floor. As the size of the inner floor area is the same as the Small room, the user's AoD distribution is expected to be high between $-20^{\circ}$ to $-30^{\circ}$ when they attract the subjects' attention. As shown in every Large condition from Fig.~\ref{fig:AoDResult}, we could observe the peak in the AoD value between $-20^{\circ}$ to $-30^{\circ}$. On the other hand, in Small size conditions, we could find the AoD peak near the AoD value between $0^{\circ}$ to $-10^{\circ}$. 

As in Study 1, the ART for non-parametric factorial ANOVA analysis ($\alpha = 0.05$) was used to conduct the multivariate analysis within the within-subject factor of \textit{Layout}, the between-subject factor of \textit{Size}, and the factor of \textit{AoD Median}. We also found significant main effect of \textit{Size} ($F(1,98) =$ 65.109, $p<$ .001), but no main effect of \textit{Layout} ($F(3,98) =$ .115, $p=$ .951) nor \textit{Size$\times$Layout} interaction effect were found ($F(3,98) =$ .525, $p=$ .666). Therefore, it was revealed that the user's AoD values in Large and Small room size was significantly different ($p<$ .001). This indicates that subjects in the small virtual room were more likely to face the front wall while walking along the path. Moreover, Fig.~\ref{fig:AoDResult} (B) shows the AoD peak in the Large size condition was presented between $-20^{\circ}$ to $-30^{\circ}$, which implies the boundary of an inner floor draws attention from subjects.

\subsubsection{Gaze Heatmap}

\begin{figure*}[tb]
 \centering
 \includegraphics[width=\textwidth]{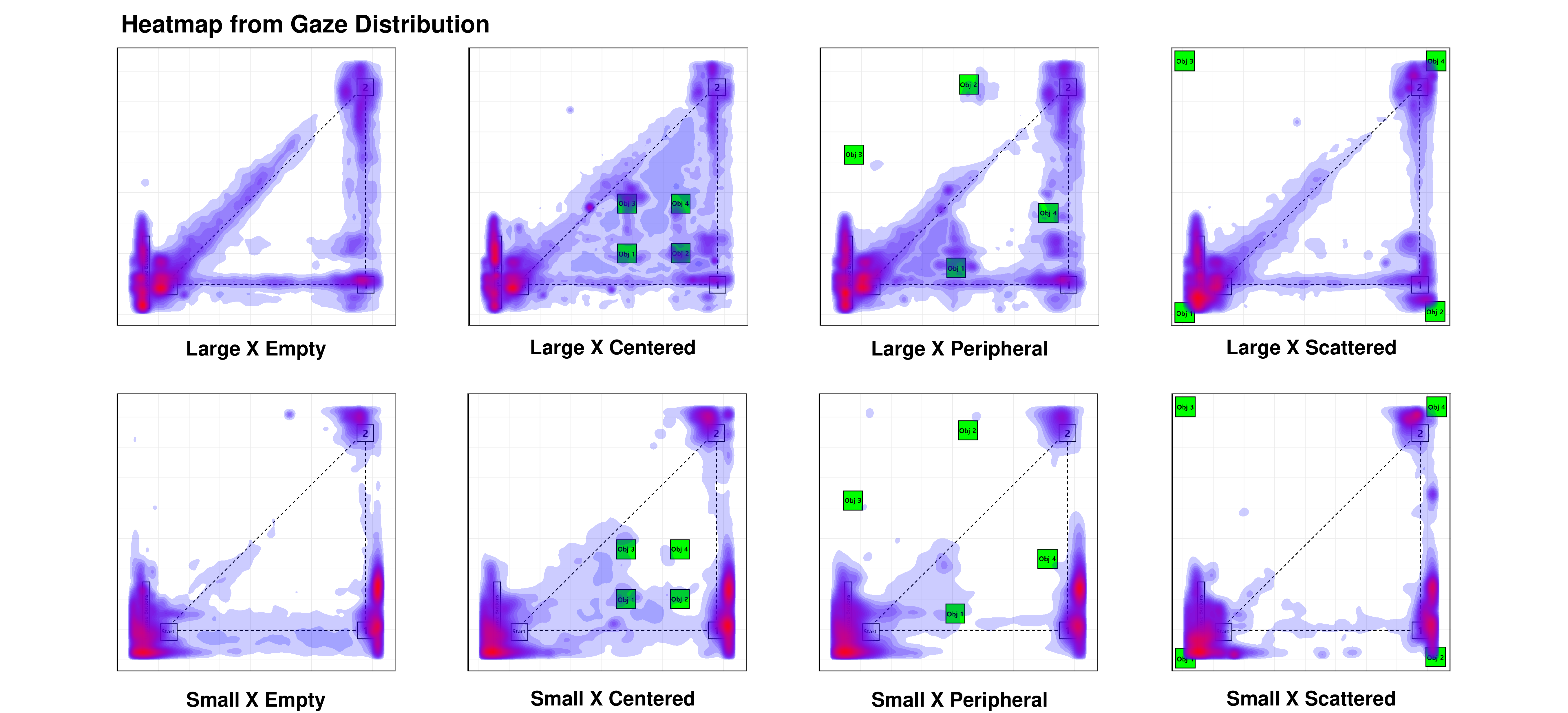}
 \caption{The top-view of the heatmap from subjects' gaze distribution in eight conditions from Study 2. Sequential paths are drawn as dashed lines and placed objects are represented as colored square boxes. Gazed data outside the inner floor was excluded to focus on the effect of spatial layout configured by objects.}
 \label{fig:Heatmap}
\end{figure*}

We utilized the gaze heatmap to observe the effect of spatial layout on the user's distribution of attention. We normalized the individual gaze information of the subjects for each condition and accumulated every subject's gaze data from the same experiment condition. While subjects walked the same path during the experiment, most of them gazed at the end of the direction in which they walked. In the four layout conditions of the Small room, the average gaze proportion reaching the wall was 76.08\% (STD = 4.18). In the Large size conditions, the average gaze proportion to the wall, the outer floor area, and the other floor areas, excluding the inner tiled floor, was 45.52\% (STD = 4.5). Fig.~\ref{fig:Heatmap} is the gaze heatmap in the eight spatial configurations using the gaze data of Study 2. Since the purpose of the gaze heatmap is to observe the user's gaze distribution changes by spatial layout, only the gaze data within the inner area 5cm away from each side of the inner floor was used. 

Fig.~\ref{fig:Heatmap} shows that the gaze heatmaps from the same spatial layout share a similar pattern regardless of the size. Results show that participants frequently focused their gaze on their walking paths. According to the spatial layout, the gaze distributions in the following pairs were similar: Empty and Peripheral,  Centered and Scattered. In the Empty and Peripheral layouts, participants mostly gazed at the paths they walked on. In addition, the Peripheral condition's gaze heatmap showed that the object located at the left top corner, which is less relevant to the path, rarely played a role in attracting the user's attention. In the Centered and Scattered layouts, on the other hand, subjects viewed not only the path but also the objects inside the path. In the Centered layout, the ratio of subjects looking at the objects inside the path is more significant. From the gaze heatmap of the Scattered layouts, two object clusters inside the path were frequently gazed and also two objects outside the path were gazed at by the subjects. In conclusion, we verified that the subjects frequently perceived objects placed in the Centered and Scattered conditions through a gaze heatmap.

\subsection{Post VR Sickness Comparison}

In Study 1, we compared the post-SSQ Total Score (TS) from two Large size conditions to determine the effect of RTG on users' VR simulator sickness~\citep{kim2022effects}. Results show there was no significant difference in the mean post-SSQ TS scores for  Large $\times$ Empty (with RTGs) and Large $\times$ Empty (no RTGs), nor between the Large $\times$ Furnished (with RTGs) and the Large $\times$ Furnished (no RTGs). Through statistical results from Study 1, we confirmed that relative translation gains are not a significant factor in VR simulator sickness. In Study 2, we compared post-SSQ TS from each condition to compare the effect of spatial layout on the VR simulator sickness. The ART for non-parametric factorial ANOVA analysis was also used to handle multifactor analysis. Results show all factors and their interaction effect showed no significant main effects (\textit{Size}: $F(1,105) =$ 1.072, $p=$ .303; \textit{Layout}: $F(3,105) =$ .711, $p=$ .548; \textit{Size} $\times$ \textit{Layout}: $F(3,105) =$ .132, $p=$ .941). We could confirm there was no significant effect of spatial layout on the VR simulator sickness.

\section{Discussion}
\label{sec:5}

\subsection{Analysis}
\label{sec:5.1}

We evaluate our three hypotheses for the study based on results from Section~\ref{sec:4}. Our first hypothesis about the impact of size on the RTG threshold was supported in Study 1, as there was a significant difference between the Large size condition and the Medium, Small one. However, our hypothesis about the impact of size on the RTG threshold was not supported, as there was no significant difference was found between the Large size condition and the Small one in Study 2. Different results from the two studies for size seem to be related to the number of conditions with distractors among the conditions experienced by the subjects. In Study 1, the Empty space without any distractors accounted for half of the experimental conditions. On the other hand, for Study 2, the proportion of the Empty room was only 25\%, and the inner floor was expressed with bright tiles, which distribute the user's attention as shown in Fig.~\ref{fig:Heatmap}. Based on these statistical results and gaze distributions from two studies, we confirmed that the effect of size in the Empty room without distractors was significant: When the space is empty, the cognitive threshold increases to a greater degree when its size is more extensive. However, the effect of the size of the space is relatively small in conditions where distractors exist.

Fig.~\ref{fig:AoDResult} (A) shows that subjects from Study 1 often looked at the virtual horizon in the Empty condition regardless of the room size. However, subjects did not perceived virtual horizon frequently in the Furnished conditions, so the impact of size is relatively small. Furthermore, Fig.~\ref{fig:AoDResult} (B) also shows that users from Study 2 did not perceive the virtual horizon frequently in every condition. This supports that the RTG threshold was affected by the size of the virtual room only if it's empty and the virtual horizon is well noticeable to subjects. However, in the case of a virtual room with distractors, the effect of room size on the RTG threshold range is less pronounced because the objects in the VR room attract users' attention.

Next, we verified the effect of the spatial layout on the RTG threshold range. The second hypothesis about the impact of spatial layout on the RTG threshold was partially accepted. The statistical results from Study 2 confirmed that the probability of ``faster'' responses in the Scattered layout is significantly lower than in the Centered layout. In addition, in the interaction effect analysis, we confirmed that the probability of ``faster'' responses in the Large $\times$ Scattered condition was significantly lower than in the Small $\times$ Centered condition. Moreover, the probability of ``faster'' responses in the Large $\times$ Peripheral condition was lower than in the Small $\times$ Empty condition. Summarizing these statistical results and estimated RTG threshold ranges from Table~\ref{tab:threshold1} and Table~\ref{tab:threshold2}, we conclude that subjects become less sensitive to an increase in their walking speed in the following order of layout conditions: Scattered, Peripheral, Empty, and Centered.

Fig.~\ref{fig:Heatmap} shows that subjects looked more at the area around the object under the Centered and Scattered conditions than the Empty and Peripheral conditions. Moreover, subjects gazed at the object more in the Centered layout than in the Scattered layout. In the post-experimental interview, many subjects said that in the Centered condition, they estimated their walking speed by referring to the speed at which objects in the space approached them. This indicates that users' regarded objects differently according to the spatial layout. Users were less sensitive to being faster when objects distract their gaze, but in the Centered layout, they utilize objects as a guide to help them notice their speed to be faster. For instance, when objects are placed at the center and gathered as a single cluster, subjects are more aware of being faster in the virtual space by using objects as reference targets to estimate their speed. On the other hand, when objects are placed scattered and come in and out of human's sight repeatedly, objects act as distractors and make users less sensitive to be faster.


Our last hypothesis, which postulated that the AoD distribution would be lower in the small-sized conditions than in larger ones, was rejected for Study 1 and Study 2. The statistical results from Study 1 show that the AoD medians from the Large conditions were significantly lower than those of Medium and Small conditions. Likewise, in Study 2, the AoD medians in Large conditions were significantly lower than those in Small conditions. These results are reflected in Fig.~\ref{fig:AoDResult}, which shows that in every Large-sized condition, subjects most frequently gazed at the area that fell between the AoD values from $-30^{\circ}$ to $-20^{\circ}$, where the inner floor's boundary exists. In every Small-sized condition, however, the area subjects gazed at most often was between $-10^{\circ}$ to $0^{\circ}$ of their AoD, meaning that they tended to look at the walls mostly while they were walking.

Fig.~\ref{fig:AoDResult} confirms that the relationship between AoD and the virtual horizon is well revealed only in Empty conditions. This means that when objects exist in a virtual space, AoD distribution is less relevant in determining the value of RTG thresholds. This is because even though the main reason for measuring AoD is to determine how often users perceive a virtual horizon, the existence of distractors distributes users' gaze and makes it challenging to find the relationship between the virtual horizon and the RTG threshold range. Rather, it is more effective to utilize the gaze heatmap for analyzing the entire space in conditions with distractors, as they could show how much each object attracts the user's gaze, as shown in Fig.~\ref{fig:Heatmap}.

\subsection{Virtual Space Rescaling Guidelines with Relative Translation Gains}

Based on Section~\ref{sec:5.1}, we propose virtual space rescaling guidelines with Relative Translation Gains (RTGs) for system developers to increase the range of adjustable movable space with RTG as follows.

\textbf{1) Increase the adjustable movable space by placing distractors in the virtual room:} 

Both study results show that the existence of distractors in the virtual room should be considered first in adjusting the virtual space through RTGs. We observed that the RTG threshold range is greater in the Small $\times$ Furnished condition than in the Large $\times$ Empty from Study 1 in Table~\ref{tab:threshold1}. This indicates that when objects are placed in a small virtual space and function as distractors, users will be less sensitive to increases in walking speed compared to a larger and empty virtual room. Moreover, the estimated RTG threshold range of Empty conditions in Study 2 was relatively higher than that of Study 1. This is because the floor tiles used in Study 2 to distinguish the inner floor area distracted the users. This means that the existence of a distractor in the virtual scene can create a virtual space that is larger than the users' actual movable space. In this study, we could find the virtual room with distractor conditions such as placing objects densely, placing objects as a scattered layout if it is not dense, or displaying a movable inner floor with bright tiles. A simple way for developers to increase the user's explorable area with distractors would be to visualize spatial boundaries with different colors or patterns on the floor.

\textbf{2) For empty spaces, set the perceived movable space to be larger than the adjusted movable space:} 

The size of the virtual room is essential for an empty virtual room. Users are less sensitive to being faster in the virtual room with a higher virtual horizon when they can recognize the horizon well while walking. They could recognize the virtual horizon well in an empty virtual room, so the threshold range of RTG shifted toward the size of the virtual room. From the perspective of space rescaling, we could increase a tracked space more when we fit it to the larger target virtual space. On the other hand, we could decrease the tracked space more when we rescaled it for matching to the smaller target virtual space. As the goal of applying RTGs to RDW is to maximize the aligned area between the virtual space and the real space, considering the effect of room size on RTG thresholds is helpful to rescale and sync the tracked real space to the target virtual space. When developers generate a new empty virtual space, the space expansion effect can be maximized if the perceivable movable space is larger than the adjusted movable space.

\textbf{3) Avoid placing objects where they are constantly in a user's sight:}

Study 2 revealed that the role of objects in a VR space differs according to the spatial layout they configured. When objects are scattered in the virtual room, they function as distractors and increase the adjustable movable space when RTGs are applied. On the other hand, when objects gather to form a single cluster, they act as aids that help users estimate their walking speed, thereby increasing the movable space. Our results show that the virtual movable space can be expanded with RTGs in the following order: Centered, Empty, Peripheral, and Scattered. In the case of Scattered and Peripheral layouts, objects frequently enter and exit the users' field of view while they are walking and constantly distract their attention from their walking speed. In the Centered layout, however, users can consistently perceive and predict their location regardless of their position and direction along the walking path. Their fixed presence in sight, along with the users' perception of how fast they approach or move away from them, help users estimate their walking speed. Therefore, it is necessary to distinguish the spatial layout of virtual space and figure out the role of objects to adaptively set the RTG threshold range and rescale the virtual space.

\subsection{Limitations}

There are some limitations in this work that should be addressed in the future. The first limitation is that the studies were only conducted in a static environment. In various locomotion-based VR applications, virtual objects and information frequently move around or pop up in the air. As this study confirmed that distractors in fixed positions in virtual space significantly influence the RTG threshold range, further investigation on whether dynamic objects or information also function as distractors in specific situations and how they affect the RTG threshold range should be conducted.

Second, we should conduct a user study with the virtual space rescaled with RTGs. As RTGs were proposed to transform the virtual space while maintaining a coordinate system between physical and virtual space, we should investigate the user experience in rescaled virtual space. Furthermore, to determine whether applying RTGs to RDW is beneficial for real-world scenarios in merging physical spaces with the target virtual space, we should compare the rescaled space based on RTGs with spaces generated from previous RDW methods.

Lastly, our study is limited to a situation where only one user occupies a virtual space. When a user co-exists with game characters or remote users in VR, various modes of interaction with them can influence the user's visual perception and recognition of changes in their walking speed. As users' gaze distribution in VR will be affected by how other users behave in the same space, factoring in the presence of others should be a needed step in expanding the scope of our current work. In particular, further studies are needed in multi-user remote collaboration scenarios where users from heterogeneous spaces gather together and communicate with each other in the mutual space.

\section{Conclusion and Future Work}

In this study, we analyzed how spatial configurations of the virtual space affect the Relative Translation Gains (RTGs) threshold range and derived guidelines on how the virtual space can be adaptively rescaled with RTG-based RDW. To verify the influence on the user's visual perception according to the composition of the virtual space, we utilized the distribution of AoDs and gaze heatmaps generated from the user's gaze data. Our results show that the three components of the virtual space--size, the existence of the objects, and the spatial layout--have a significant effect on the RTG threshold. The adjustable movable space can be increased further when distractors exist in the virtual room. Moreover, the adjustable movable space can be expanded when the perceived movable space is larger than the actual movable space in empty VR environments. Lastly, the adjustable movable space increases in a scattered layout where objects function as distractors, but it decreases in a centered layout where objects function as guides by which users can estimate their walking speed.

When applying translation gain-based RDW, proper threshold values must be set to avoid VR simulator sickness. Using the virtual space rescaling guidelines derived from our results, developers can determine proper RTG threshold ranges according to the configuration of the target virtual space. Our proposed translation gain-based spatial deformation method can synchronize the target virtual space with the tracked physical space while maximizing the movable area for the user. More specifically, this can be achieved by precisely matching the boundaries of physical objects to virtual ones through RTG-based deformation, which can unify the coordinate system of the two spaces.

Our future work will explore the effect of dynamic objects and information on the RTG threshold range. In addition, we will determine the performance of RTG-based rescaling methods with 3D indoor scene data sets obtained from tools such as Matterport 3D~\citep{chang2017matterport3d} and compare the generated explorable virtual space results with previous methods~\citep{lehment2014creating, keshavarzi2020optimization, keshavarzi2022mutual}. We will also experiment with a scenario where multiple VR clients' avatars move around in the same target VR space according to their adaptive RTG values. Ultimately, we aim to develop and apply RTG-based RDW to generate a mutual collaborative space for Extended Reality (XR) that aligns heterogeneous physical spaces to a single VR space with a unified coordinate system.

\section{Declarations}

\subsection{Funding}

This research was supported by National Research Council of Science and Technology (NST) funded by the Ministry of Science and ICT (MSIT), Republic of Korea (Grant No. CRC 21011) and the National Research Foundation of Korea (NRF) grant funded by the Korea government (MSIT) (No. 2021R1A2C2011459).

\subsection{Competing interests}
The authors have no relevant financial or non-financial interests to disclose.


\bibliography{sn-bibliography}


\end{document}